# ИЗУЧЕНИЕ МЕТОДА СТАТИСТИЧЕСКИХ ИСПЫТАНИЙ В ПЕДАГОГИЧЕСКОМ ВУЗЕ

**Майер Р.В.**, д.п.н., доцент
ФГБОУ ВПО "Глазовский государственный педагогический институт"
г. Глазов

Аннотация. В статье анализируется проблема изучения метода статистических испытаний на занятиях по компьютерному моделированию в педагогическом вузе. Предлагаемая методика предусматривает имитационное моделирование различных стохастических процессов. Среди них передача информации по каналу связи, колебания маятника в потоке воздуха, отклонения альфа–частиц атомами золота, образование перколяционного кластера и т.д.

## A STUDY OF MONTE-CARLO METHOD IN A TEACHERS' TRAINING INSTITUTE
**Mayer R.V.**

Annotation. The problem of Monte-Carlo method study at computer simulation lessons in a Teachers' Training Institute is reviewed in the article. The suggested technique envisages the simulation modelling of various stochastic processes. They include transmission of information via a communication link, oscillation of a pendulum in an air stream, deflection of alpha particles by Au atoms, formation of a percolating cluster, etc.

Для изучения систем, функционирование которых не определяется полностью их параметрами, начальным состоянием и внешними воздействиями, а зависит от каких–то случайных факторов, используется метод статистического моделирования (или метод статистических испытаний). Он состоит в многократном проведении испытаний с последующей статистической обработкой получающихся результатов. Этот подход позволяет исследовать целый класс стохастических процессов, среди которых: передача сообщений по каналу связи, обучение вероятностного автомата, поведение дискретно–детерминированных систем, на вход которых поступают случайные сигналы, функционирование систем массового обслуживания и т.д. Метод статистических испытаний — один из важнейших методов исследования стохастических систем, поэтому его следует изучать в курсе компьютерного моделирования в педагогическом вузе. В 4 и 5 главах электронного учебника "Компьютерное моделирование" [4, 5] предложена методика такого изучения, предусматривающая построение дискретных и непрывных компьютерных моделей, изучение различных систем массового обслуживания, вероятностных автоматов, вероятностных клеточных автоматов и т.д. Рассмотрим несколько задач, анализ которых позволяет понять сущность метода статистических испытаний [1–6].

**<u>Задача 1.</u>** Промоделируйте одномерные случайные блуждания молекулы газа, используя фибоначчиевый генератор случайных чисел. Полу-чите распределение конечной координаты $z_N$ блуждающей точки при различном количестве совершенных шагов $N$. Убедитесь в том, что квадрат смещения случайной величины $z_N$ прямо пропорционален $N$. Решение этой задачи можно найти в электронной книге [3].

**Задача 2. Нитяной маятник, состоящий из подвешенного на нити тела, находится в горизонтальном потоке воздуха. Скорость движения воздуха в потоке изменяется случайным образом, а направление остается неизменным. Необходимо изучить движения маятника, получить кривую распределения его угловой координаты $\varphi$, найти ее среднее значение и среднеквадратическое отклонение (СКО).** Решение этой задачи рассмотрено в [5]. На рис. 1.1 изображен график зависимости угла отклонения маятника $\varphi$ от времени, а на рис. 1.2 — распределение значений угла $\varphi$.

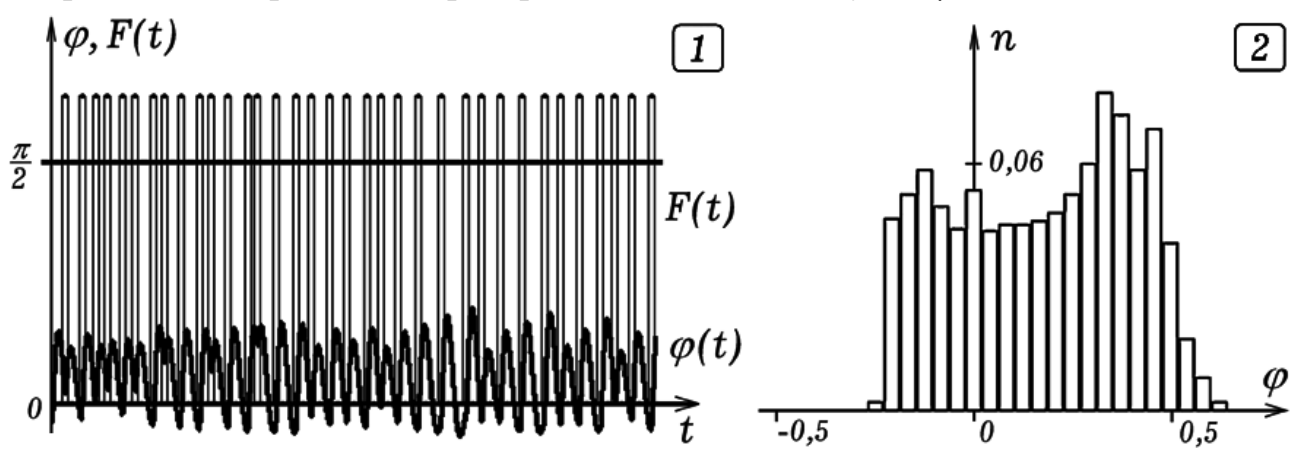

Рис. 1. Изучение случайных колебаний маятника в потоке воздуха.

**Задача 3. Создайте имитационную модель опыта Резерфорда по рассеянию альфа–частиц атомами золота. Рассчитайте траекторию движения частицы в поле отталкивания двух атомов. Методом статистических испытаний изучите зависимость числа альфа–частиц от угла отклонения при случайных значениях прицельного параметра.** Решение представлено в [5]. На рис. 2 показаны силы, действующие на альфа–частицу и распределение их угла отклонения $\alpha$, полученное методами статистического моделирования.

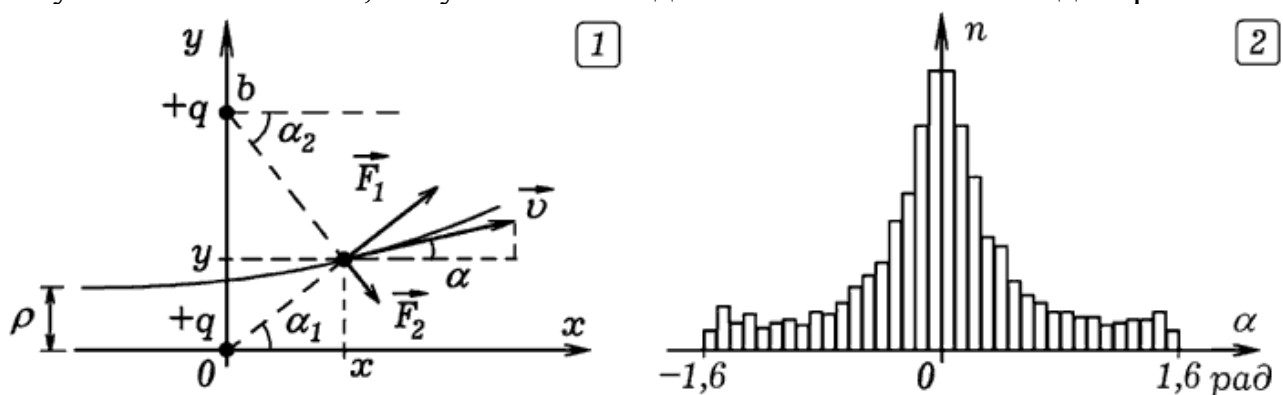

Рис. 2. Отклонение альфа–частиц в поле атома золота.

**Задача 4.** Имеется металлическая сетка $M \times M$, расположенная между двумя электродами, из которой случайным образом с заданной вероятностью $q$ удалены некоторые узлы или ячейки (а с вероятностью $p = 1 - q$ заняты). При больших $q$ будет вырезано слишком много узлов, и исчезнет путь, соединяющий нижний электрод с верхним, — сетка перестанет проводить ток. Если вырезано мало узлов, то будет существовать один или несколько перколяционных кластеров, пронизывающих данную структуру насквозь и соединяющих электроды. Необходимо изучить зависимость вероятности $P$ образования перколяционного кластера от $p$. Решение задачи приведено в [3, 4]. На рис. 5.1 ячейки, входящие в один кластер, закрашены одним цветом. Хорошо виден перколяционный кластер, пронизывающий всю структуру и соединяющий верхний и нижний электроды. Для изучения зависимости вероятности $P$ образования перколяционного кластера от вероятности $p$ наличия узла используется метод статистических испытаний. Сначала, исходя из заданной вероятности $p$ наличия занятой ячейки, случайным образом формируется ячеистая структура (рис. 5.1), после чего определяется, содержит она перколяционный кластер или нет. Эта процедура многократно повторяется, что позволяет определить вероятность перколяции $P$ при данном значении $p$. Затем проводится аналогичный вычислительный эксперимент при других $p$ и строится график зависимости $P(p)$ (рис. 5.2).

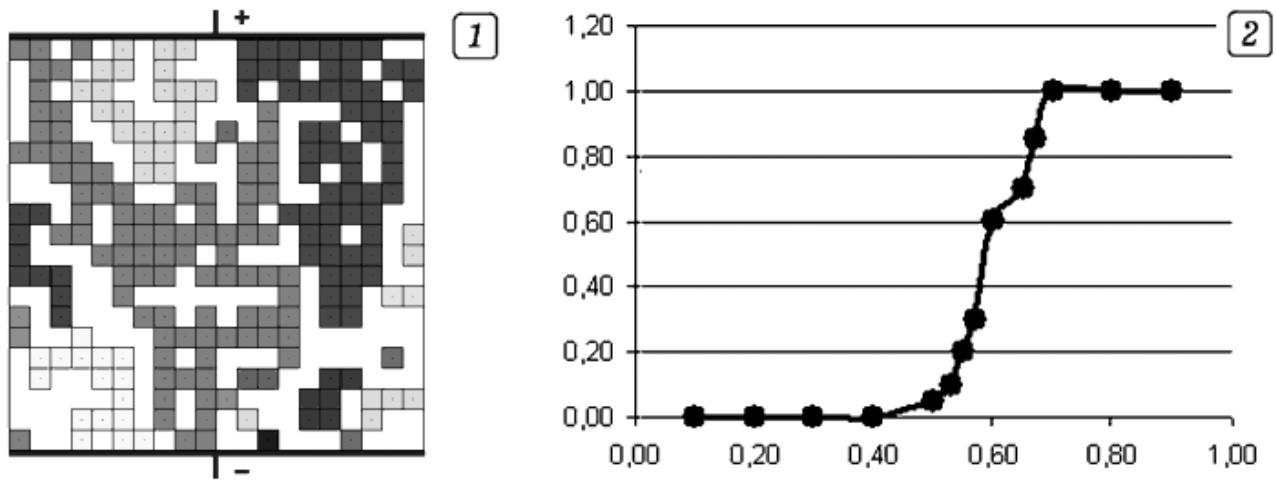

Рис. 3. Возникновение перколяционного кластера.

**Задача 5.** Технологический процесс состоит из 5 операции. Длительность и вероятность правильного выполнения каждой операции задаются матрицами: $\tau_i = \{1{,}6;\ 2{,}7;\ 1{,}4;\ 3{,}8;\ 2{,}6\}$ и $p_i = \{0{,}6;\ 0{,}7;\ 0{,}4;\ 0{,}8;\ 0{,}6\}$. Если операция выполнена неверно, то она повторяется снова. Необходимо вычислить среднее время выполнения технологического процесса и дисперсию этой величины. Решение этой задачи приведено в [4].

**Задача 6.** Источник вырабатывает сообщение 10110010…1: Кодер разбивает его на блоки длиной $D-1$ и добавляет 1 бит четности так, что получаются кадры длиной $D$. Они поступают в канал связи, в котором с вероятностью $p$ вносятся ошибки (инвертируются биты), и, пройдя через него, попадают в декодер. **На передачу 1 бита затрачивается 1 такт (допустим, 1 мс). Реализуется система с переспросом: декодер выявляет кадры с ошибками и по каналу переспроса посылает сигнал о повторе передачи соответствующего кадра. На его повторную передачу снова затрачивается N тактов. Сигнал по каналу переспроса не вносит задержки. Промоделируйте этот процесс на ПЭВМ, определите скорость передачи.** Решение задачи приведено в [4]. В программе организован цикл, в котором моделируется покадровая передача информации. При каждой итерации время $t$ увеличивается на $D$. В кадре длиной $D$ происходит ошибка с вероятностью $pD$. Генерируется случайное число $x$ и если оно меньше $pD$, то происходит ошибка и по каналу связи повторно передается тот же кадр (для этого $i$ уменьшается на 1). При безошибочной передаче кадра на экран просто выводится время $t$. Скорость передачи определяется так: $\upsilon = N_K(D-1)/t$, где $(D-1)$ — число информационных бит, $N_K$ — число кадров, $t$ — число тактов. Она зависит от длины кадров и вероятности ошибки $p$. Используемая программа представлена ниже.

```
uses crt, dos;
const Chislo_k=500; Dlina_k=8; p=0.05;
var I,t: longint; x, skorost :real;
BEGIN
    For i:=1 to Chislo_k do begin t:=t+Dlina_k;
      x:=random(1000)/1000; writeln('KADR ',i, x);
      If x<p*Dlina_k then begin
      Writeln('OSHIBKA V KADRE ',i,' ',t); i:=i-1; end;
      If x>=p*Dlina_k then writeln(t);
    end;
skorost:=Chislo_k*(Dlina_k-1)/t;
Writeln(t,' ',skorost); Readkey;
End.
```

В случае, когда длина кадра $D = 8$, и сообщение передается без помех ($p = 0$), скорость передачи полезной информации получается равной $\upsilon = 0{,}875$ бит/с. В самом деле, к семи информационным битам прибавляется восьмой проверочный, поэтому скорость передачи можно рассчитать так: $\upsilon = 7/8 = 0{,}875$ бит/с. Проведя серию вычислительных экспериментов можно убедиться, в том что при увеличении вероятности $p$ ошибки скорость $\upsilon$ передачи информации уменьшается: 1) при $p_1 = 0{,}001$, $\upsilon_1 = 0{,}867$ бит/с; 2) при $p_2 = 0{,}01$, $\upsilon_2 = 0{,}800$ бит/с; 3) при $p_3 = 0{,}02$, $\upsilon_3 = 0{,}724$ бит/с; 4) при $p_4 = 0{,}05$, $\upsilon_4 = 0{,}522$ бит/с.

<u>Задача 7</u>. **Постройте график зависимости скорости передачи информации от длины кадра, если вероятность ошибки постоянна и равна 0,02; 0,1; 0,3.** Используется та же программа. Если $p = 0{,}1$, то при длине кадра 2 или 3 бита скорость передачи невелика за счет большого числа проверочных битов четности. С ростом длины кадра она уменьшается из–за увеличения вероятности ошибки в кадре и затрат времени на повторную его передачу. Существует оптимальная длина кадра, при которой скорость максимальна. Результаты моделирования для $p = 0{,}05$ и $0{,}02$ — кривые 1 и 2 на рис. 4.1.

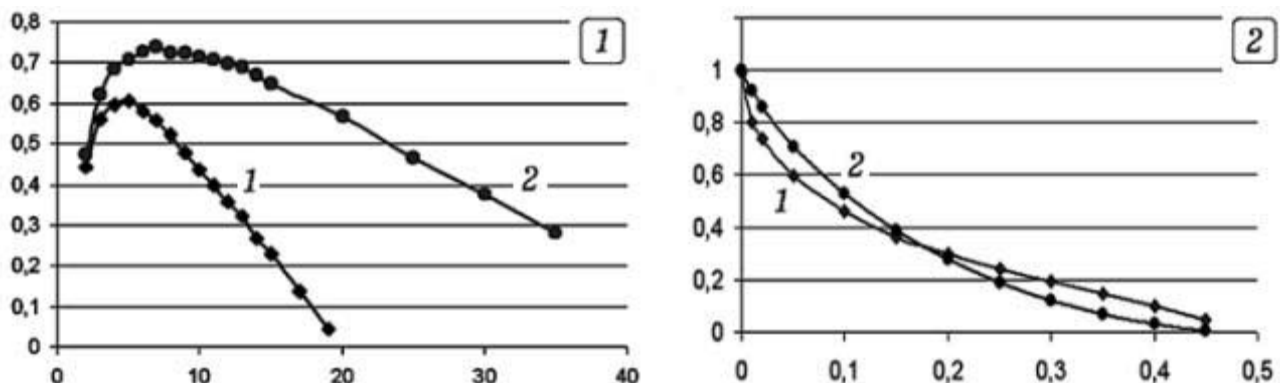

Рис. 4. Результаты моделирования передачи информации по каналу связи.

**Задача 8. Считая, что емкость канала связи C равна максимальной скорости передачи информации, изучите зависимость емкости канала от вероятности ошибки, постройте график. Сравните полученные результаты с расчетными значениями для двоичного симметричного канала с шумом.** Емкость (пропускная способность) двоичного симметричного канала связи с вероятностями ошибки $p$ и правильной передачи $(1-p)$ равна $C = C_0(1 + p\log_2 p + (1-p)\log_2(1-p))$. Для нахождения емкости $C$ моделируемого канала связи с переспросом зададим вероятность ошибки 0,05 и найдем скорости передачи при различных длинах кадра. Обнаружим, что при $N = 5$ скорость передачи максимальна и равна 0,60, — эту величину и следует приближенно считать емкостью канала связи $C$. Повторим эту процедуру при других $p$, каждый раз определяя максимальную скорость передачи. Построим график 1 зависимости емкости моделируемого канала связи от вероятности ошибки (рис. 4.2). Видно, что при увеличении p от 0 до 0,5 она уменьшается от 1 до 0. Эта кривая 1 похожа на расчетную кривую 2 для двоичного симметричного канала связи, но точного совпадения нет [3].

**Задача 9. На вход канала связи поступает последовательность символов из трехбуквенного алфавита $A = \{a_1, a_2, a_3\}$, вероятности использования которых определяются матрицей $p_i = (0{,}3;\ 0{,}25;\ 0{,}45)$. На выходе канала связи получается поток символов из алфавита $A' = \{a_1, a_2, a_3, b\}$, где $b$ — ошибочный символ. Статистические свойства канала связи задаются стохастической матрицей:**

$$P = \begin{pmatrix} 0{,}7 & 0{,}1 & 0{,}1 & 0{,}1 \\ 0{,}1 & 0{,}8 & 0{,}1 & 0{,}0 \\ 0{,}2 & 0{,}2 & 0{,}5 & 0{,}1 \end{pmatrix}.$$

**Необходимо промоделировать передачу информации по каналу связи и методом статистических испытаний определить вероятность ошибки.** Элементами матрицы являются вероятности $p_{i,j}$ появления на выходе канала связи $j$–ой буквы из алфавита $A'$, когда на вход канала связи поступает $i$–ая буквы из алфавита $A$ ($i = 1,2,3$; $j = 1,2,3,4$). Используемая программа [4] содержит цикл, в котором методом выбора по жребию определяется входной символ, а затем с помощью матрицы вероятностей $P$ разыгрывается выходной символ. При этом осуществляется подсчет относительного числа ошибок $n_{ош}/n$, и результат выводится на экран. Экспериментируя с программой, можно установить, что после 2000 – 3000 испытаний результаты вычисления $n_{ош}/n$ приобретают статистическую устойчивость.

Предлагаемая методика преподавания метода статистических испытаний, отдельные элементы которой представлены в работах [3, 4, 5], позволяет сформировать понимание сущности этого метода, способствует повышению интереса студентов к компьютерному моделированию.

## Библиографический список